\documentclass[12pt,a4paper]{iopart}

\usepackage[T1]{fontenc}
\usepackage[latin9]{inputenc}
\usepackage[english]{babel}
\usepackage{color}
\usepackage{float}

\expandafter\let\csname equation*\endcsname=\relax
\expandafter\let\csname endequation*\endcsname=\relax

\usepackage{amsmath}
\usepackage{amssymb,amsthm}

\usepackage{graphicx}
\PassOptionsToPackage{normalem}{ulem}
\usepackage{ulem}
\usepackage{babel}

\newcommand{\Lk}{Lk}
\newcommand{\Wr}{W\!r}

\begin{document}

\title{Writhe-induced knotting in a lattice polymer}

\author{E Dagrosa$^1$, A L Owczarek$^1$ and T Prellberg$^2$}
\address{$^1$ Department of Mathematics and Statistics,
  The University of Melbourne, Parkville, Vic 3010, Australia.}
\address{$^2$ School of Mathematical Sciences, Queen Mary University
  of London, Mile End Road, London E1 4NS, UK.}
\ead{e.dagrosa@student.unimelb.edu.au, owczarek@unimelb.edu.au, t.prellberg@qmul.ac.uk}

\begin{abstract}
We consider a simple lattice model of a topological phase transition in open polymers.
To be precise, we study a model of self-avoiding walks on the simple cubic lattice 
tethered to a surface and weighted by an appropriately defined writhe.
We also consider the effect of pulling the untethered end of the polymer
from the surface.

Regardless of the force we find a first-order phase transition which
we argue is a consequence of increased knotting in the lattice polymer,
rather than due to other effects such as the formation of plectonemes.
\end{abstract}

\pacs{02.50.Ng,02.70.Uu,05.10.Ln,36.20.Ey,61.41.+e,64.60.De,89.75.Da}
\ams{82B20, 82B41, 82B80}
\submitto{\JPA}
\maketitle

\section{Introduction}

Over the past years, there has been continuing interest in topological phase transitions
of polymers. One focus has been on modelling DNA.
For example, in experiments \cite{edsgcl.29795367720100101,DEUFEL_QUARTZ,ABRUPT_BUCKLING}
single molecules of twist-storing polymers such as double stranded
DNA can be held torsionally constrained and under constant stretching
force. These experiments show abrupt phase transitions known as buckling, and the formation
of conformational structures known as plectonemes.

We are interested here in what topological and/or geometrical phase transitions might occur in a single-stranded
polymer: can one, for example, find buckling and the formation of plectonemes.

To this end, we consider a model of self-avoiding walks on the simple cubic lattice
in a half-space with an appropriately defined writhe. 
We will refer to the variable conjugate to writhe as (pseudo-)torque, and work in an ensemble in which this torque
is held constant. We perform simulations with the flatPERM algorithm \cite{PrellbergFlatpermB,PrellbergFlatpermA}, in which
walks are grown from an end that is tethered to the surface. It is important to realise that in these simulations
only one end is held fixed. 

We find that upon varying torque there is a first-order phase transition between states of small and
high average writhe. However, in contrast to the experimental situation described above,
we do not see a buckling transition. Instead, the low and high writhe states are dominated by different 
distributions of effective knot types. This scenario seems unchanged by the presence of a pulling force applied to the endpoint. 
Because this transition is driven by the change of knot-type, we argue that it is also insensitive
to the choice of lattice. 

While it is not clear how to create an experimental realisation of our model, we are interested in the principle
of the existence of a topological phase transition. Another motivation comes from
recent work \cite{GENERLARIBBON}, where it was shown that the linking number of a lattice ribbon, which is a lattice version 
of a double-stranded polymer \cite{Rens_LatRib_DoubStrandPoly}, is equal to the writhe of the center line of the ribbon.
The center line is a restricted three-dimensional self-avoiding walk on the half-integer simple cubic lattice, so a model weighting 
the writhe of such a self-avoiding walk can be 
considered as weighting the linking number of a lattice ribbon. 
There are two key differences between the experiments mentioned above and such a model based on a lattice ribbon.
The experiments are conducted in a fixed linking number ensemble.
This means that the number of times the molecule is turned is equivalent to the linking number of the DNA. This relation between linking number and turns is guaranteed
by the experimental setup, which prevents the DNA from passing over its endpoints. Otherwise the DNA could just change its linking number by passing over its endpoints so that
adding turns to the molecule becomes irrelevant in the sense of statistical mechanics. Another consequence of the fact that the DNA cannot pass over its endpoints is that it cannot form knots.
In contrast, the lattice ribbon model related to the model we study here allows for knotting. Such an ensemble is not easily realized in experiments, since you would presumably still require the above relation between the number of turns and the linking number to hold.

\subsection{The Model}

Consider self-avoiding walks (SAW) on the simple
cubic lattice. An $n$-step SAW $R_{n}$ is formed by $n$ edges, or equivalently $N:=n+1$ vertices
$\omega_{i}\in\mathbb{Z}^{3},\, i=1,..,N$ such that 
\begin{enumerate}
\item $\omega_{i}\neq\omega_{j}$ $\forall\, i\neq j$, and
\item $\left\Vert \omega_{i+1}-\omega_{i}\right\Vert =1$ $\forall\, i<N$.
\end{enumerate}
We refer to $n$ as the length of this SAW.

We anchor the SAWs at the origin $\omega_{1}=\left(0,0,0\right)$
and restrict them to lie in the positive half-plane $\left(\omega_{i}\right)_{z}\geq0$.
The pulling force is aligned along the $z$ direction so that the extension
of the SAW is given by $h=\left(\omega_{N}\right)_{z}$. Let $C_{n,w,h}$
be the number of SAWs of length $n$ with extension $h$ and
a parameter $w$ which corresponds to  writhe and which is defined below. Then, the canonical
partition function reads
\begin{equation}
Z_{n}\left(F,\, T\right)=\sum_{h}\sum_{w}C_{n,w,h}\,\exp\left[F\cdot h+T\cdot w\right].\label{eq:Partition-Function}
\end{equation}

The parameter $T$ shall be referred to as (reduced) torque, whereas $F$ shall be referred to as (reduced) force.  

Technically, writhe is defined only for a closed curve. To make sense
of the writhe of an open curve, it is usually necessary to close the
open curve in a well defined way. 
We approach this problem in a different way.
The writhe of a self-avoiding closed
curve or polygon $P$ \cite{SUMNERS_WRITHES} on the simple cubic lattice can be expressed as
\begin{equation}
\Wr\left(P\right)=\frac{1}{4}\sum_{j=1}^{4}\Lk\left(P,\, P+\sigma_{j}\right),
\end{equation}
where $P+\sigma_{j}$ is a copy of $P$ translated by a vector $\sigma_{j}$
and $\Lk$ is the linking number of the two polygons $P$ and $P+\sigma_{j}$. The vectors $\sigma_{j}$ can be
chosen to be $\sigma_{1}=\left(0.5,\,0.5,\,0.5\right)$, $\sigma_{2}=\left(-0.5,\,0.5,\,0.5\right)$,
 $\sigma_{3}=\left(-0.5,\,-0.5,\,0.5\right)$, $\sigma_{4}=\left(0.5,\,-0.5,\,0.5\right)$.
On the other hand, linking number can be computed as the number of
signed crossings in a projection plane. Suppose the curves $P$ and
$P+\sigma_{j}$ are projected into the $xy$-plane at $z=-\infty$, then
the operator $\hat{S}_{xy}$ can be defined to sum up all the signed crossing
$\epsilon\left(c\right)$ 
\begin{equation}
\hat{S}_{xy}\left(P,\, P+\sigma_{j}\right)=\frac{1}{2}\sum_{\text{$c$ is crossing}}\epsilon\left(c\right),
\end{equation}
so that $\Lk=\hat{S}_{xy}(P, P+\sigma)$ and in particular $\hat{S}_{xy}=\hat{S}_{yz}=\hat{S}_{zx}$.
When the curves are not closed, linking number is not defined but
the operators $\hat{S}$ remain defined. While the operator
result in general depends on the projection plane, by averaging
over all planes the result becomes trivially independent of the plane.
For an open self-avoiding curve $R_{\text{sc}}$, we define the integer
\begin{equation}
w\left(R_{\text{sc}}\right)=\sum_{j=1}^{4}\left[\hat{S}_{xy}\left(R_{sc},\, R_{\text{sc}}+\sigma_{j}\right)+\hat{S}_{yz}\left(R_{\text{sc}},\, R_{\text{sc}}+\sigma_{j}\right)+\hat{S}_{zx}\left(R_{\text{sc}},\, R_{\text{sc}}+\sigma_{j}\right)\right].
\end{equation}

When the curve is closed then $w\left(P\right)=12\, \Wr\left(P\right)$,
so that $w\left(R_{\text{sc}}\right)$ is closely related to the writhe.
It can be considered an approximation of the writhe of a closed curve $P[R_{\text{sc}}]$
that is obtained when one tries to close a given open self-avoiding curve $R_{\text{sc}}$ ``simply''.
$w\left(R\right)$ is invariant under rotations and translations that
respect the lattice symmetry. Thus, $w$ is a true microcanonical
parameter of a SAW on the simple cubic lattice. Under reflections at a coordinate
plane, $w$ picks up a sign. The quantity $w$ shall be referred to as the writhe of
the walk. 

\subsection{Knots}

An embedding of the circle $S\rightarrow\mathbb{R}^{3}$ is called
a knot \cite{1107.216220110711}. Any knot defines an equivalence
class called the knot type $K$. Two knots are equivalent if one knot
can be transformed into the other via homotopy transformation. There
are two kind of knot types. Prime knots like the unknot $0_{1}$ and
composite knots like the concatenation of two trefoils $\left(3_{1}\right)\#\left(3_{1}\right)$.

For any SAW $R_{n}=\left\{ \omega_{i}\right\} _{i=1,..,N}$ as defined
above, define the corresponding knot by the following procedure. In
the first step, add $N$ vertices $\omega_{k}^{'}$ $(k\in 1,..,N)$ in $z$ direction to the end of the SAW.
The coordinates of these vertices read $\left(\omega_{k}^{'}\right)_{x}=\left(\omega_{N}\right)_{x}$,
$\left(\omega_{k}^{'}\right)_{y}=\left(\omega_{N}\right)_{y}$,
$\left(\omega_{k}^{'}\right)_{z}=\left(\omega_{N}\right)_{z}+k$.
Then, add $2\, N$ vertices in $x$ direction. Then, add vertices in $-z$ direction until the $z$-component becomes $-1$.
Add vertices in $\pm y$ direction until
the $y$ component becomes zero. Finally, add   vertices in $-x$ direction
until the $x$ component becomes zero. The last added vertex and the first
vertex $\omega_{1}=\left(0,0,0\right)$ of the SAW are adjacent. Connecting
them forms a reference lattice polygon $P^{\text{ref}}\left(R_{n}\right)$.
When the lattice polygon is self-avoiding, it is a knot with knot
type $K$. Define the knot type of the walk $R_{n}$ to be
\begin{equation}
K_{R}\left[R_{n}\right]=K\left[P^{\text{ref}}\left(R_{n}\right)\right].
\end{equation}
When the reference polygon is not self-avoiding, the knot type of the
walk shall be called undefined. 
While the choice of the reference polygon is not unique, we expect the gross features of our conclusion not to be affected,
as they relate to changes in whether dominant configurations in the ensemble are typically knotted and the way we construct our
reference polygon is consistently applied.

For any SAW $R_{n}\in K_{R}$ with defined knot type, we can define
its writhe by the writhe of its reference polygon
\begin{equation}
\Wr\left(R_{n}\right):=\Wr\left(P^{\text{ref}}\left(R_{n}\right)\right).
\end{equation}
In particular, one may compare $w\left(R_{n}\right)$ to $12\, \Wr\left(R_{n}\right)$. 

We can also define the expectation value of a knot type $K$ for polygons
$P$ on the simple cubic lattice as 
\begin{equation}
\left\langle K\right\rangle =\frac{\sum_{P}\delta\left(P\in K\right)x\left(P\right)}{\sum_{P}x\left(P\right)},\label{eq:Knot-type-expectation}
\end{equation}
where $x\left(P\right)$ is a Boltzmann weight with respect to some microcanonical parameters. 

We use an algorithm similar to the one used in \cite{cond-mat/070378420070329,4472730720091021,1979122520060214}
to detect whether the reference polygon of a SAW is the unknot.

\section{Algorithm and Data}

We use the flatPERM algorithm \cite{PrellbergFlatpermB,PrellbergFlatpermA}
to produce estimates for the numbers $C_{n,w,h}$.
The flatPERM algorithm grows a SAW one vertex at a time by selecting
randomly one out of $a_{n}$ possibilities that keep the walk self-avoiding.
The larger $a_{n}$, the more valuable is the selection, so that
a SAW grown into unoccupied regions have a higher statistical weight
\begin{equation}
W_{n}=\prod_{k=0}^{n-1}a_{k}.
\end{equation}

Let $S$ be the number of started growth chains, then estimates of
$C_{n,w,h}$ are given by
\begin{equation}
C_{n,w,h}^{(\text{est})}=\frac{1}{S}\sum_{i}W_{n}^{\left(i\right)}\delta\left(w\left(R_{n}\right)-w\right)\,\delta\left(h\left(R_{n}\right)-h\right)\,.
\end{equation}
The flatPERM algorithm uses local pruning and enrichment to enable uniform sampling.
Let $R_{n}$ be a SAW of length $n$ with writhe $w$ and extension $h$. When the ratio 
\begin{equation}
r=\frac{W_{n}\left(R_{n}\right)}{C_{n,w,h}^{(\text{est})}}\label{eq:r}
\end{equation}
is larger than $1$ the walk is enriched. Otherwise the walk becomes
pruned. Enriching is done by making $c=\min\left(\left[r\right],\, a_{n}\right)$
copies of the walk and setting their weights to $\frac{1}{c}W_{n}$.
Each copy is grown into a different direction, so that the the weights
of the walks of length $n+1$ will be given by 
$W_{n+1}^{1}=W_{n}\,a_{n},\ldots,W_{n+1}^{c}=W_{n}\left(a_{n}-c+1\right)$. 
On the other hand, pruning consists in continuing to grow the walk
with probability $r$ and setting its weight to $C_{n,w,h}^{(\text{est})}$.
Therefore, growing of the walk is discontinued with probability $(1-r)$.
The estimates $C_{n,w,h}^{(\text{est})}$ are believed \cite{PrellbergFlatpermB,PrellbergFlatpermA} to converge towards
their true values with the number of started growth chains $S$. A better measure of
the effective sample size in a simulation, that has proven to be useful in practice,
is provided by keeping track of the quantity
\begin{equation}
S{}_{n,w,h}^{(\text{eff})}=\frac{1}{n}\sum_{i}n_{\text{ind}}^{(i)}\label{eq:Seff},
\end{equation}
where $n_{\text{ind}}^{\left(i\right)}$ is the number of vertices grown
independently. These are the number of steps since the walk with parameters
$\left(n,w,h\right)$ was last enriched. By changing the ratio
in Equation~(\ref{eq:r}) to
\begin{equation}
r=\frac{W_{n}\left(R_{n}\right)}{C_{n,w,h}^{(\text{est})}}\frac{S}{S{}_{n,w,h}^{(\text{eff})}}\;,
\end{equation}
the local effective sample size $S{}_{n,w,h}^{(\text{eff})}$ is taken into account for pruning and enrichment.

We compute estimators $\left\langle Q\right\rangle _{n}^{est}$ for observables $Q$ using
\begin{equation}
\left\langle Q\right\rangle _{n}^{est}\left(T,\, F\right)=\frac{\sum_{h,w}C_{n,w,h}^{(\text{est})}Q\, \exp\left[F\cdot h+T\cdot w\right]}{\sum_{h,w}C_{n,w,h}^{(\text{est})}\exp\left[F\cdot h+T\cdot w\right]}.
\end{equation}
If instead of estimating $C_{n,w,h}$ one is interested in estimating partition functions for the constant-force ensemble,
one can run one-parameter simulations at fixed force $F$ by applying the same algorithm to $Z_{n,w}(F)=\sum C_{n,w,h}\exp[F\cdot h]$ instead.

We use logarithmic coding \cite{berg-Log-Eq20} to cope with large
numbers. We gather data for $\left|w\right|$ rather than $w$ and
assume by symmetry that $C_{n,-w,h}=C_{n,w,h}$.

Suppose $K$ independent simulations were performed and let $S_{k}$ with $k=1,...,K$ denote the number of
growth chains completed in the $k$-th simulation. We obtain a statistical
average 
\begin{equation}
\overline{Q}_{n}=\frac{1}{K}\sum_{k}\left\langle Q\right\rangle _{n}^{est\,\left(k\right)},
\end{equation}
and estimate the standard error 
\begin{equation}
S\!E=\sqrt{\frac{\sum S_{k}^{2}}{\left(\sum S_{k}\right)^{2}}}\sqrt{\frac{\sum_{k}S_{k}\left(\left\langle Q\right\rangle _{n}^{est\,\left(k\right)}-\overline{Q}_{n}\right)^{2}}{\sum S_{k}}}.
\end{equation}
Assuming that a sample of $K$ simulations yields $\overline{Q}_{n}$ that
are normal distributed around the true value $\left\langle Q_{n}\right\rangle $,
the probability that the interval $\overline{Q}_{n}\pm1.96\, S\!E$ covers
$\left\langle Q_{n}\right\rangle $ is $95\%$.

\subsubsection*{Data}

\begin{figure}[t!]
\begin{center}\includegraphics[width=4in, keepaspectratio]{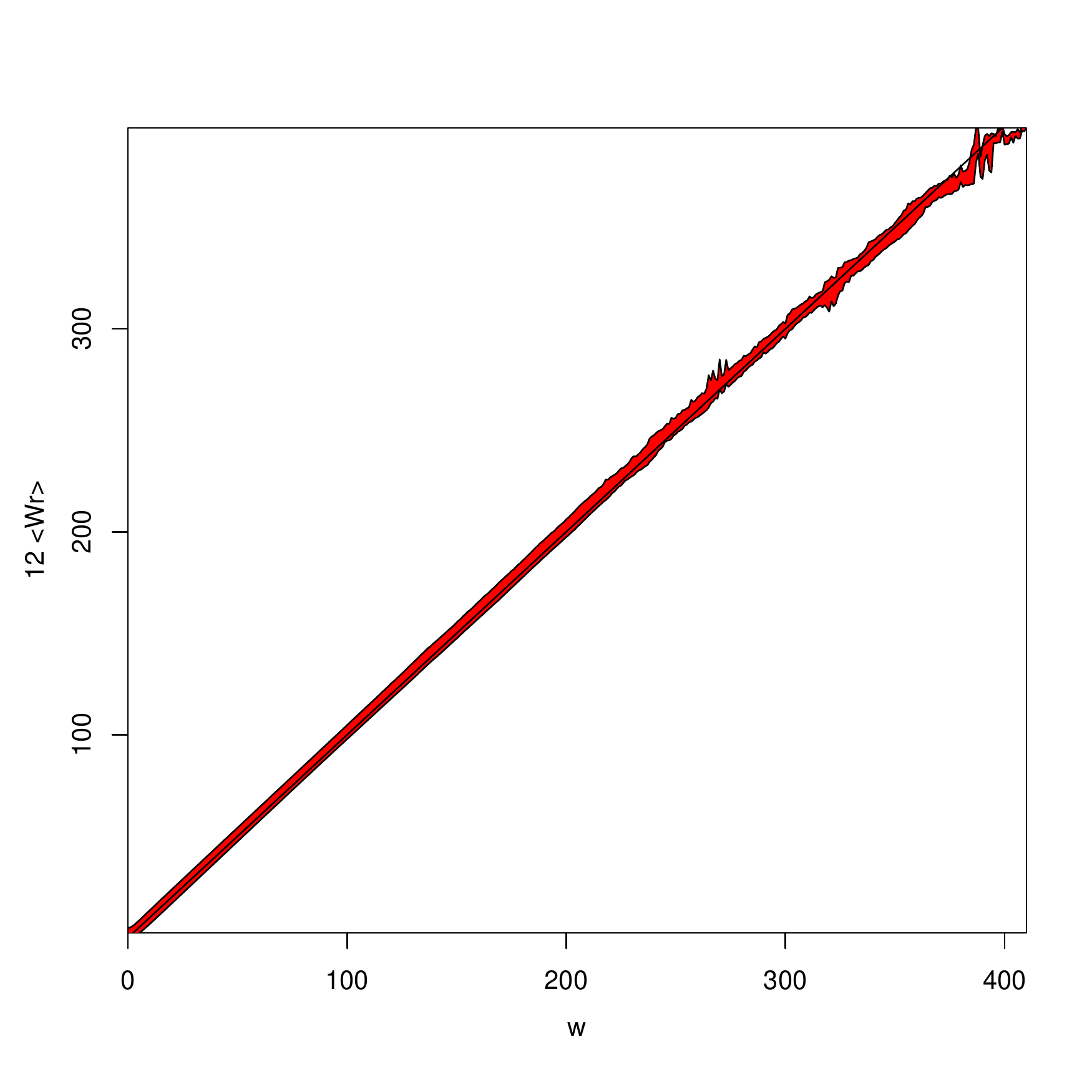}\end{center}
\caption{\label{fig:N120-Writhe}
Shown is the average writhe $12\,\left\langle \Wr\right\rangle $ of a reference polygon for walks with $n=120$ steps and fixed writhe $w$. The width of the curve indicates the standard error. The diagonal (black line) is shown for comparison.}
\end{figure}

We ran eight independent flatPERM simulations up to length $n=120$
to collect data for $C_{n,w,h}$. The total number of growth chains
for these simulations is $S_{total}\approx4.6\times10^{5}$. The shortest
simulation has $S_{\text{short}}\approx0.7\times10^{4}$, the longest $S_{\text{long}}\approx7.7\times10^{4}$
growth chains. The combined number of produced samples at length $120$
is $N_{\text{samples}}\approx1.0\times10^{11}$ of which $N_{\text{eff}}\approx1.1\times10^{9}$
can be regarded as effectively independent in the sense of Equation~(\ref{eq:Seff}).

To check the consistency of our approach with regards to using our definition of the writhe of the walk $w$, we have compared it to the writhe $\Wr$ 
of a reference polygon for those configurations for which a reference polygon exists. Figure~\ref{fig:N120-Writhe} shows the results taken
from the simulation at $F=0$. Within error bars, there is excellent agreement between both quantities.

In addition we ran eight one-parameter flatPERM simulations with
maximum length 200 to collect data for $Z_{n,w}\left(F=0\right)$.
The attributes of these simulations read $S_{total}  \approx  6,1\times10^{6}$, $S_{\text{short}}  \approx  5.0\times10^{5}$, $S_{\text{long}}  \approx 1.0\times10^{6}$, $N_{\text{samples}}  \approx  5.4\times10^{10}$ and $N_{\text{eff}}  \approx  4.5\times10^{8}$. 

Another four one-parameter flatPERM simulations with up to  
$N=256$ vertices were run to collect data for $Z_{n,w}\left(F=\log(1.4)\right)$. Each with $5.0\times10^{5}$ growth chains, so that $S_{total}  =  2\times10^{6}$, $N_{\text{samples}}  =  2.4\times10^{10}$ and $S_{\text{eff}}  \approx  1.5\times10^{8}$. 

\section{Results}

\begin{figure}[t!]
\centering\includegraphics[scale=0.5]{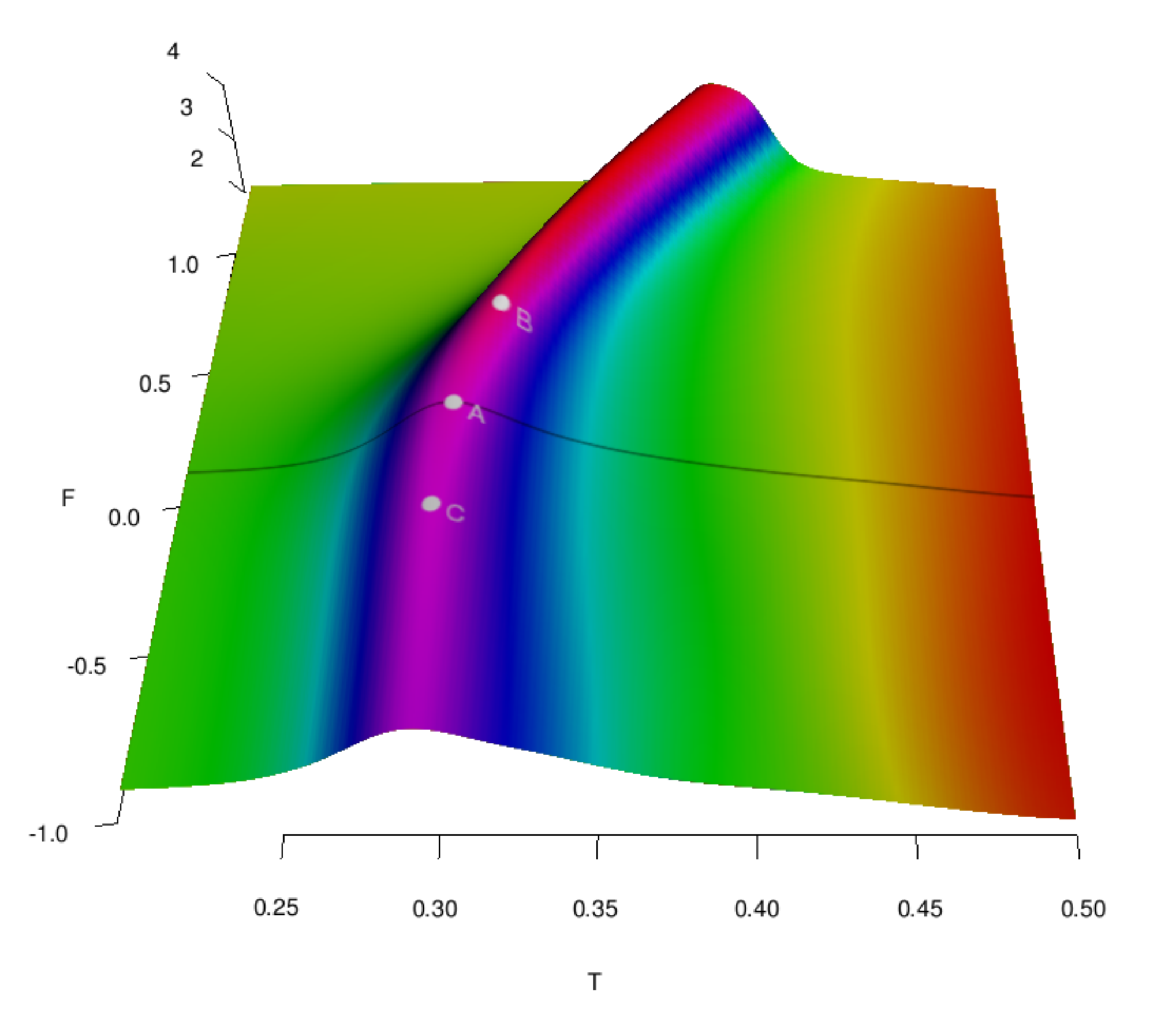}
\caption{\label{fig:Pseudo-Phase-diagram}Maximal eigenvalue $\lambda$ of the matrix of second derivatives at length $n=120$.
The $z$-axis shows $\log_{10}\left(\lambda\right)$. Torque $T$ is in $x$
direction, force $F$ in $y$ direction. The marked points are $A=\left(0.298,\,0\right)$,~$B=\left(0.312,\, \log(1.4)\right)$,
$C=\left(0.294,\,-\log\left(1.4\right)\right)$.}
\end{figure}

We start by considering the data produced by the full two-parameter simulations. Figure~\ref{fig:Pseudo-Phase-diagram} shows the maximal eigenvalue
$\lambda$ of the matrix of second derivatives of $\log\, Z_{n}\left(T,\, F\right)$
for walks of length $n=120$. The line of peaks indicates the possibility of fluctuations that diverge as the system size increases, and hence
the possibility of a phase transition. Hence this line separates regions of low and high torque. 

\begin{figure}[t]
\centering\includegraphics{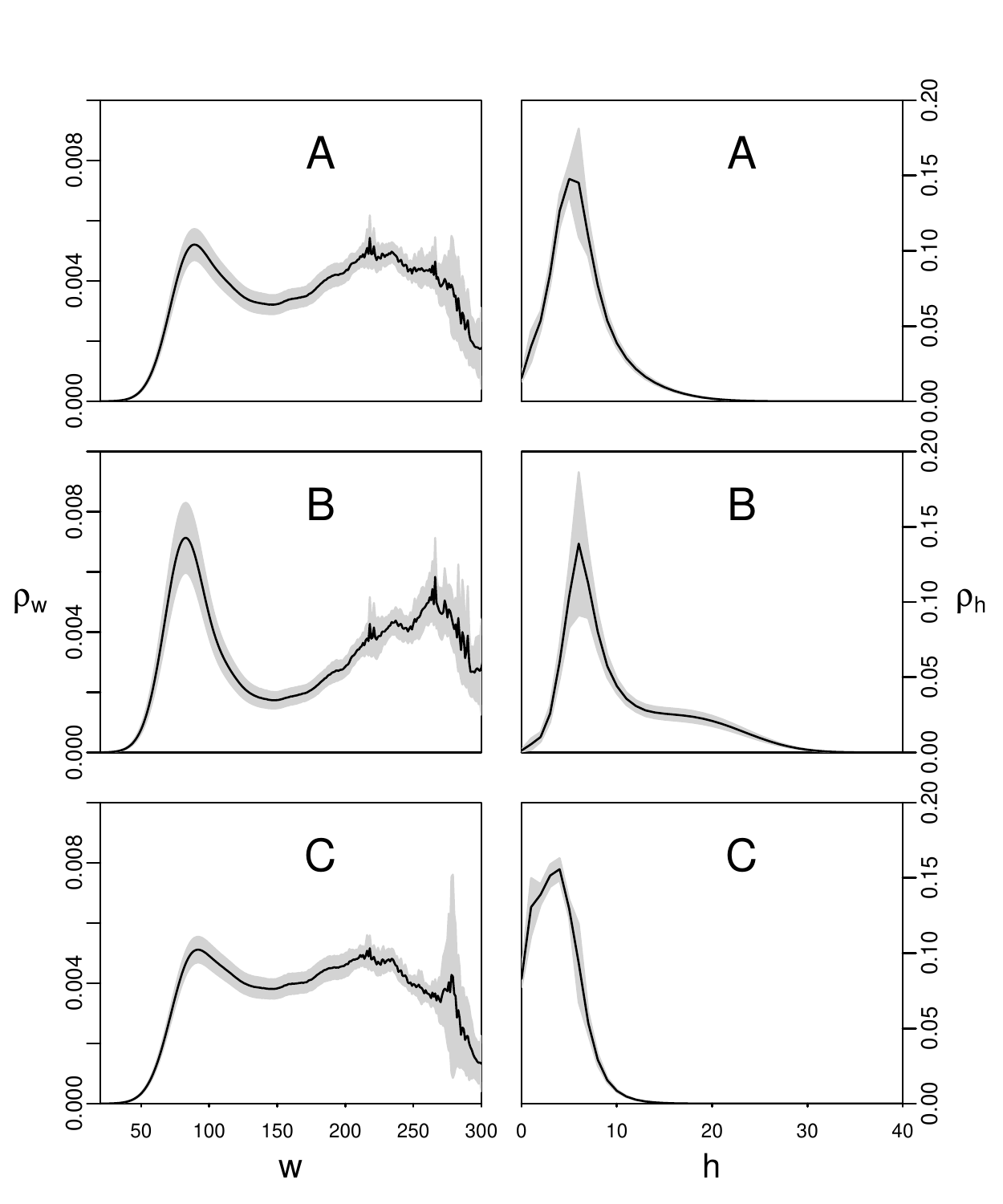}
\caption{\label{fig:ProbDistN120}Distributions for length $n=120$ at the points
$A$, $B$, $C$. The left hand side shows the writhe distribution,
the right hand side shows the distribution of the extension. At integer values of $w$ ($h$) a vertical slice of the gray shaded area (at integer values)  corresponds to the confidence interval.}
\end{figure}

\begin{figure}[t]
\includegraphics{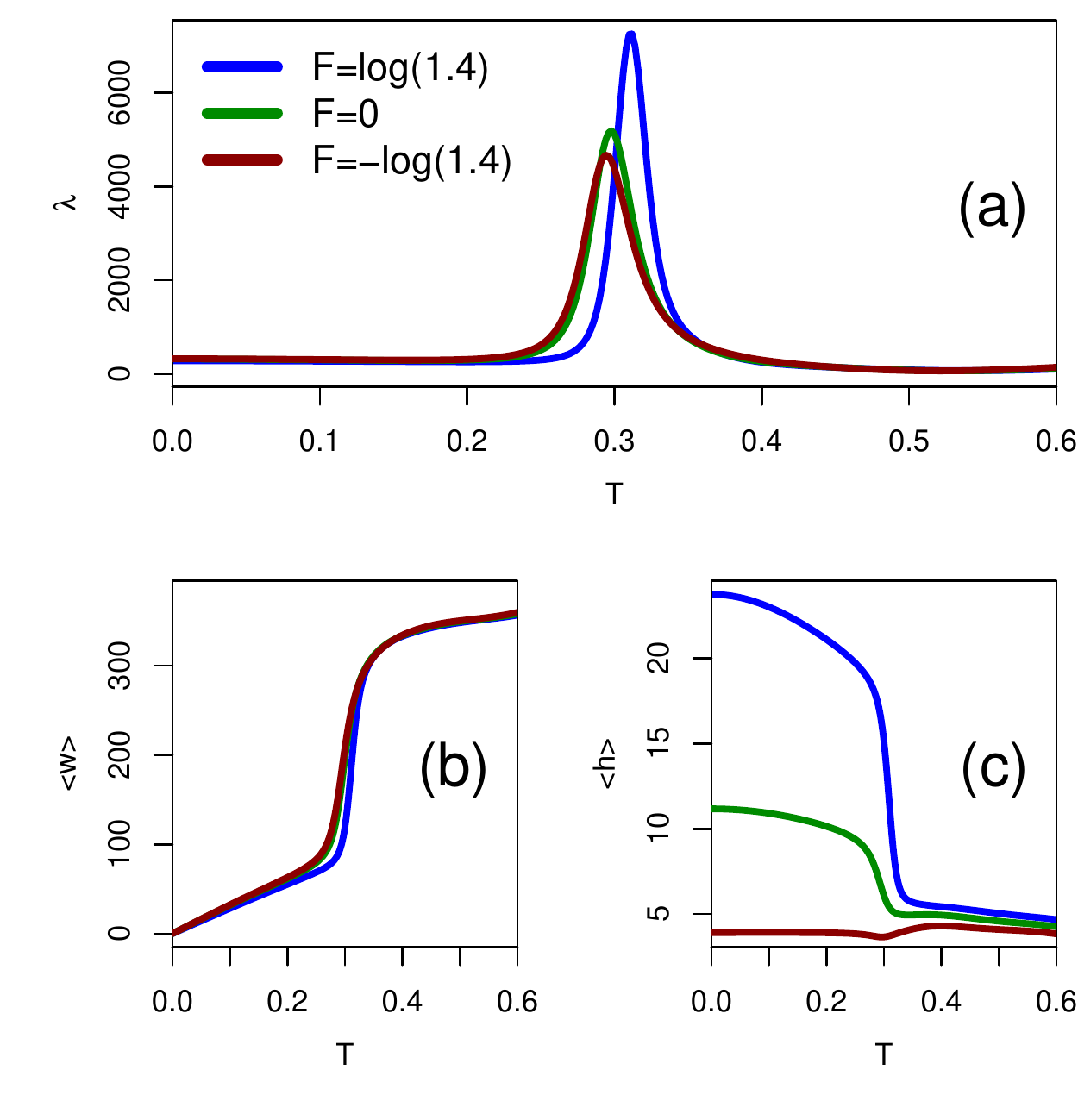}
\caption{\label{fig:N120-Observables}Observables $\lambda$ (a), $\left\langle w\right\rangle $ (b)
and $\left\langle h\right\rangle $ (c) at different pulling forces
for walks of length $n=120$. Confidence intervals are not shown. The position of the transition moves  towards increasing values of $T$ with the pulling force.}
\end{figure}

To investigate the possibility of a phase transition, we have studied three particular
values of the stretching force $F=0,\,\pm\log\left(1.4\right)$.
The locations of the maximum value of fluctuations at these particular values are indicated by the three
points marked  in Figure~\ref{fig:Pseudo-Phase-diagram}. 
The probability distributions
of $w$ and $h$ are shown in Figure~\ref{fig:ProbDistN120}. While
the distribution of the extension appears to stay unimodal, the 
writhe distributions are bi-modal, which is indicative of a first-order phase transition.

Figure~\ref{fig:N120-Observables}
shows the behaviour of $\lambda$, $\left\langle w\right\rangle $
and $\left\langle h\right\rangle $ on lines of constant stretching
force $F=0,\,\pm\log\left(1.4\right)$. Figure~\ref{fig:N120-Observables}~(a)
shows that the fluctuations become highly peaked at the three points A, B, and C. 
The average writhe takes on a profile expected for a first-order transition, with
a sharp jump at the same points, as seen in Figure~\ref{fig:N120-Observables}~(b).
On the other hand, the average extension, which is shown in Figure~\ref{fig:N120-Observables}~(c),
changes in a smoother fashion, and only when the force is non-negative. It is effectively constant
when the force is negative. We conclude that any transition is related to a sharp change in the writhe
and a possible divergence in the fluctuations of the writhe.

\begin{figure}[t]
\includegraphics{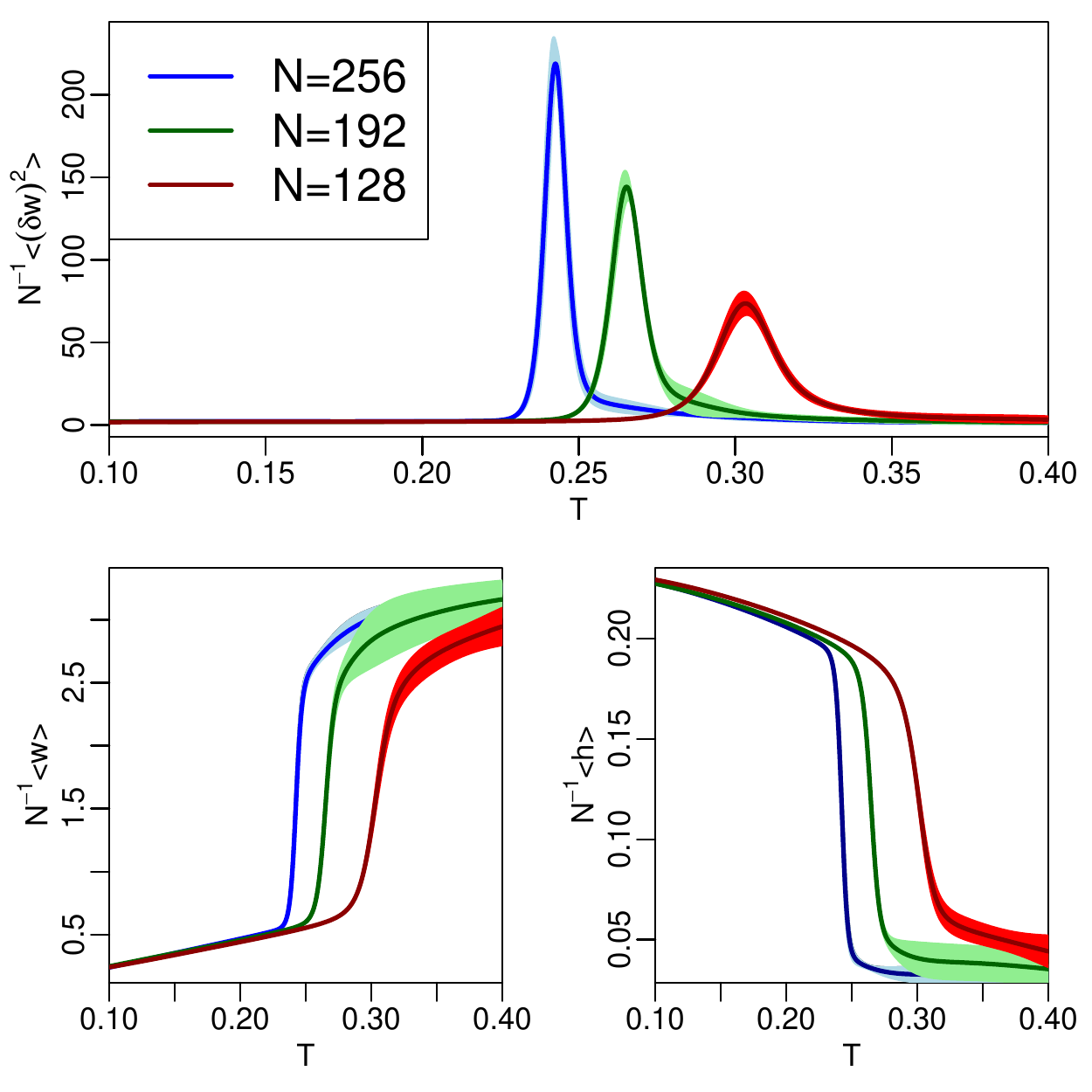}
\caption{\label{fig:Observables} The observables $N^{-1}\left\langle \left(w-\left\langle w\right\rangle \right)^{2}\right\rangle $,
$N^{-1}\left\langle w\right\rangle $, $N^{-1}\left\langle h\right\rangle $, at
lengths $N=256,\,128,\,196$ and $F=\log\left(1.4\right)$. The shaded areas represent confidence intervals.  }
\end{figure}

To establish whether there is a true phase transition, we consider the scaling of the observables 
$  N^{-1}\left\langle \delta w\right\rangle \equiv N^{-1}\left\langle \left(w-\left\langle w\right\rangle \right)^{2}\right\rangle $,
$N^{-1}\left\langle w\right\rangle $ and $N^{-1}\left\langle h\right\rangle $
with length. We first look at positive force $F=\log\left(1.4\right)$ for walks with up to $256$ vertices. 
From Figure~\ref{fig:Observables} it is evident that the jump in the writhe and extension become sharper with increasing length, and that
the height of the peak of the fluctuations in writhe also sharply increases in length.
One can note that the height of the peak at $N=256$ is more than double the height of the peak at $N=128$. Such strong increase is again indicative of a first-order phase transition.

\begin{figure}[t]
\includegraphics{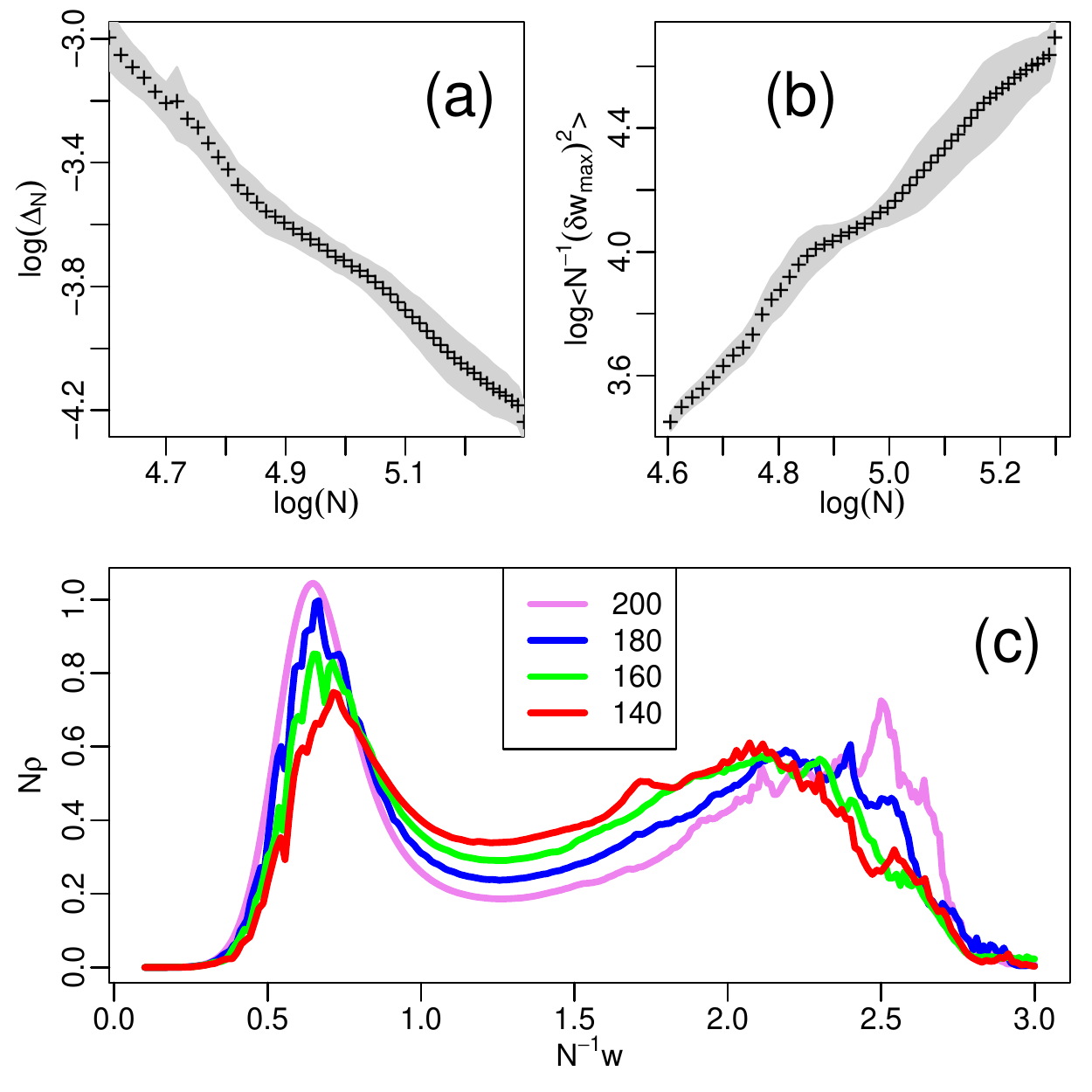}
\caption{\label{fig:Scaling} (a) Scaling of the half widths and (b) the maximum
in writhe fluctuation for $n=100$ to $n=200$. The gray shaded area is obtained by connecting the $95\%$ confidence intervals.
(c) Scaling of the writhe distribution. The bridge between the peaks
deepens with the length. The corresponding torque values of maximal
writhe fluctuation $\langle(\delta w_{max})^2\rangle=\langle(\delta w)^2\rangle(T_n^{*})$ are $T_{200}^{*}=0.2442$, $T_{180}^{*}=0.2545$, $T_{160}^{*}=0.2662$, and $T_{140}^{*}=0.2799$.}
\end{figure}

Turning to the case when no pulling force is applied ($F=0$), we consider the scaling of the half-width of the peak of the writhe $\Delta_n$, the peak height of the writhe $\delta w_{\text{max}}$, and the writhe distribution itself. 
Figure~\ref{fig:Scaling}~(a) shows that the half-width $\Delta_n$ decreases to zero faster than $1/n$, and Figure~\ref{fig:Scaling}~(b) shows that the peak height $\delta w_{\text{max}}$ increases faster than linear in $n$. This is again indicative of a first-order transition, as the build-up of a bi-modality goes along with stronger super-linear scaling.
Figure~\ref{fig:Scaling}~(c) shows the scaling of the writhe distribution at the points $T_{n}^{*}$ of maximum peak height $\delta w_{\text{max}}$.
We find a bi-modal distribution, as expected for first order transitions, with the gap between the two peaks becoming more pronounced as the length of the walk increases.

To obtain a good estimate of the critical temparature, we use a standard scaling Ansatz \cite{Rensburg_MC_AW} that involves 
two pairs of lengths $n_1$/$n_2$ and $m_1$/$m_2$. At the critical value of the torque $T=T_{crit}$ 
\begin{equation}
\frac{\log(\langle w\rangle_{m_1}/\langle w\rangle_{m_2})(T)}{\log(m_1/m_2)}=\frac{\log(\langle w\rangle_{n_1}/\langle w\rangle_{n_2})(T)}{\log(n_1/n_2)}
\end{equation}  	 
should hold. Using the choices $n_1=190$, $n_2=170$, $m_1=160$, $m_2=144$, we find $T_{crit}\approx 0.22$.

We conclude that there exists a first-order phase transition between phases of low and high average writhe with a sharp jump in the writhe per unit length.
We now turn our attention to understanding the difference between these two phases in more detail.

\begin{figure}[t]
\centering\includegraphics[width=4in, keepaspectratio]{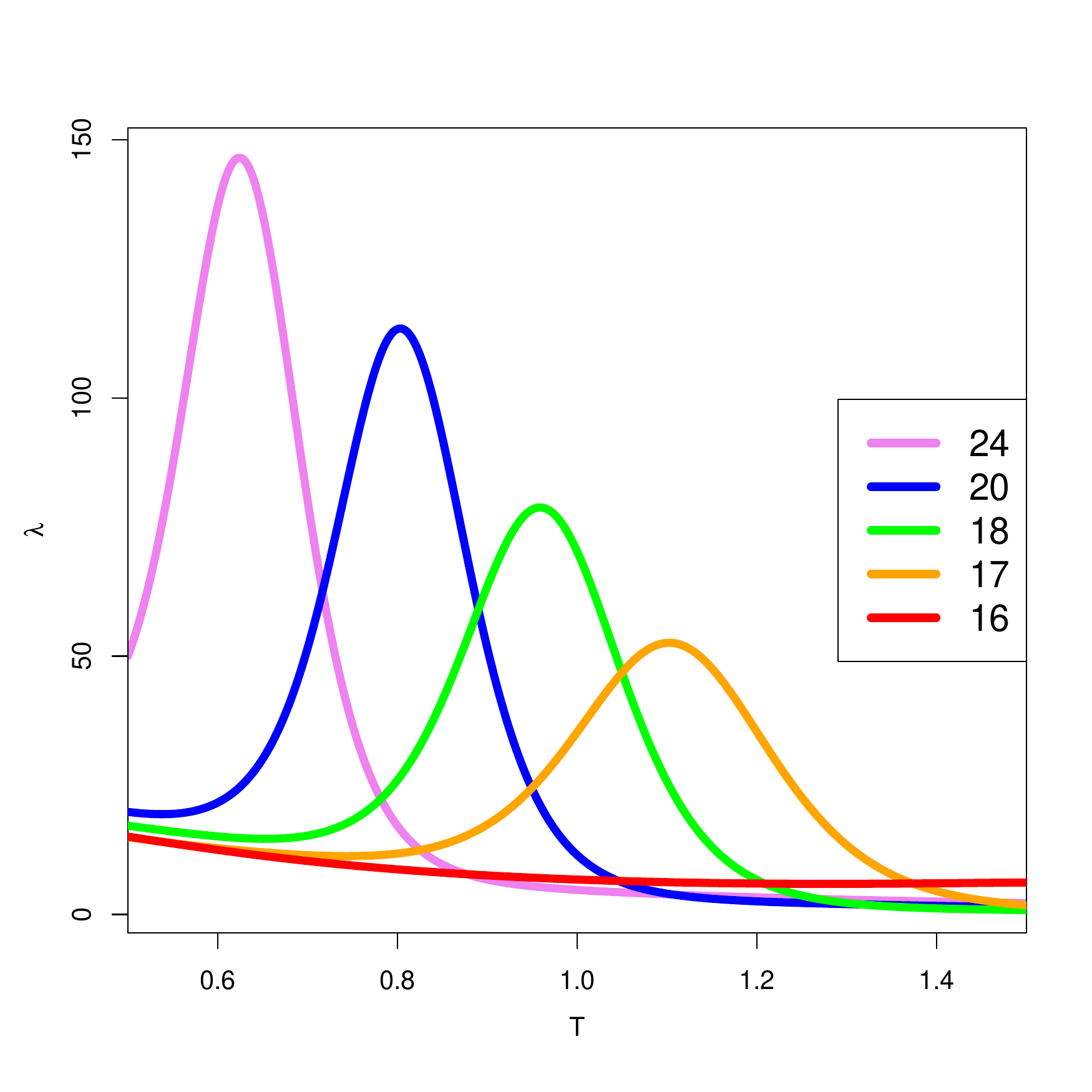}
\caption{\label{fig:lambda_N16_N24} The graphic shows the maximum eigenvalue
of the matrix of second derivatives at $F=0$ for different lengths
$n$ of the walk. Note the peak is absent at length $n=16$. The peak
appears abruptly at length $17$. The minimal length for the reference
knot to be other than the unknot is $15$. However, $17$ is the minimal
length for the SAW to pass through the formed loop and gain maximum
writhe. The peak rapidly moves with $n$ towards smaller values of
the torque and becomes more pronounced.}
\end{figure}

Figure~\ref{fig:lambda_N16_N24}
shows $\lambda\left(F=0,\, T\right)$ for SAWs between $n=16$ and $24$. When $n=16$, there is no peak in $\lambda$ on varying the torque, while for 
$n>16$ there exists a distinct peak in $\lambda$. This can be understood by noting that $n=17$ is the minimal length for a SAW to form a loop and pass through
that loop with exactly one edge. This allows the SAW to gain a significant
amount of writhe. When we look at the reference polygon of a walk
that has stepped through a loop we find that the polygon is knotted. The knot that is produced is the trefoil $3_1$, and so the open configuration forms a
partial trefoil. This indicates that the existence of knots is important in driving this transition.

\begin{figure}[t]
\centering\includegraphics[width=6in,keepaspectratio]{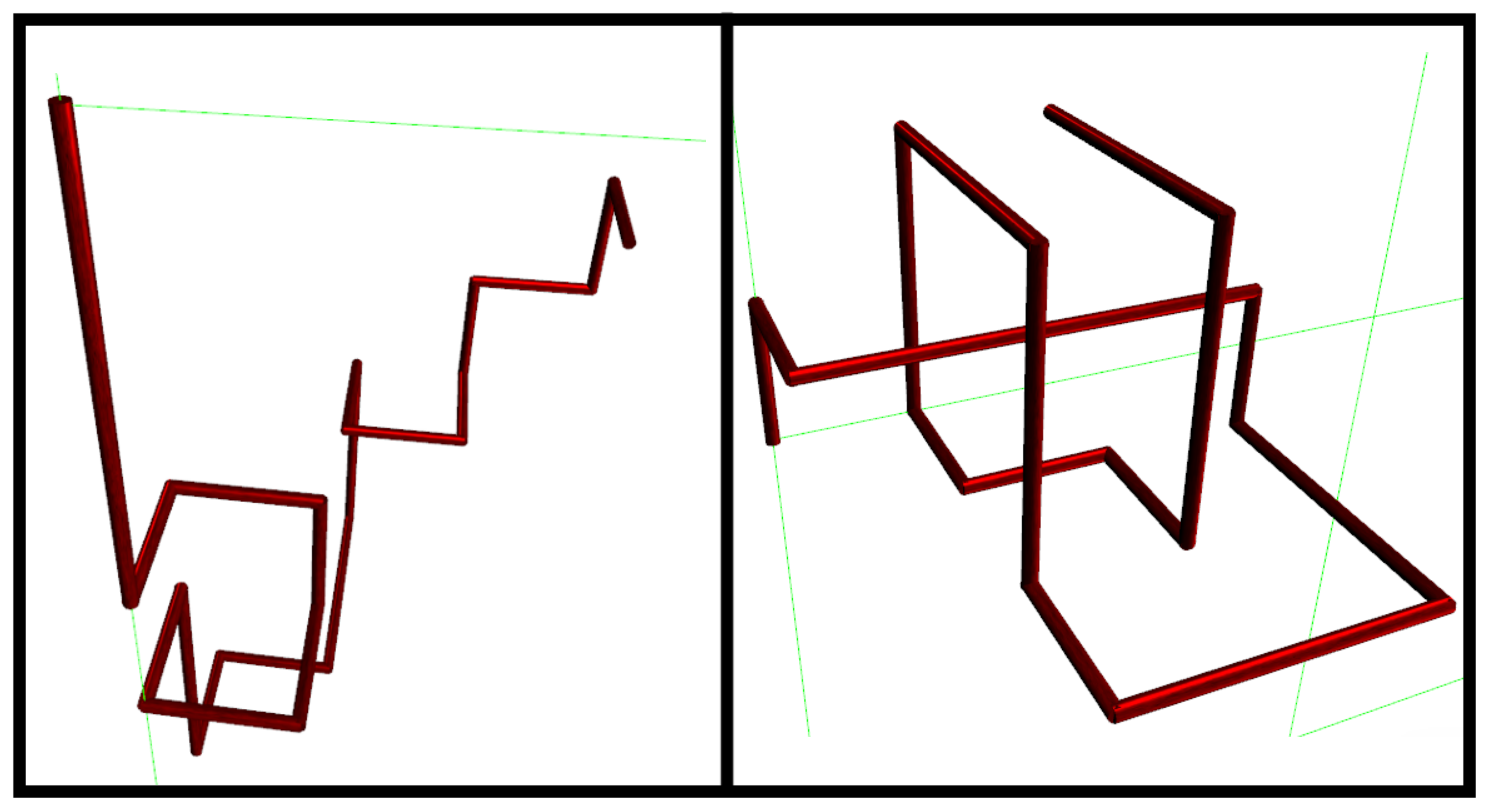}
\caption{\label{fig:Equilibrium}The graphic shows representative configurations
of the two phases for a SAW of length $n=24$ on both sides of the phase transition at $T=0.452$ and $T=0.796$.
At these two values of the torque the fluctuation in writhe is half of the peak fluctuation.
The low-torque (left) and high torque (right) configurations have writhe $w=18$ and $w=45$, respectively. These values are
close to the expectation values at these torques. The $z$ direction is to the right, so that by forming the reference polygon
we find that the given configurations with $w=18$ and $w=45$ are equivalent to the unknot and the trefoil, respectively.}
\end{figure}

Figure~\ref{fig:Equilibrium} shows two typical configurations of $24$-step SAWs. On the left-hand side, a SAW with writhe $w=18$ is shown, which will occur for small values
of torque. The reference polygon for this configuration is the unknot. On the right-hand side, a SAW with writhe $w=45$ is shown, which will occur for large values of torque. The reference polygon for this configuration is the trefoil.

\begin{figure}[t]
\centering\includegraphics{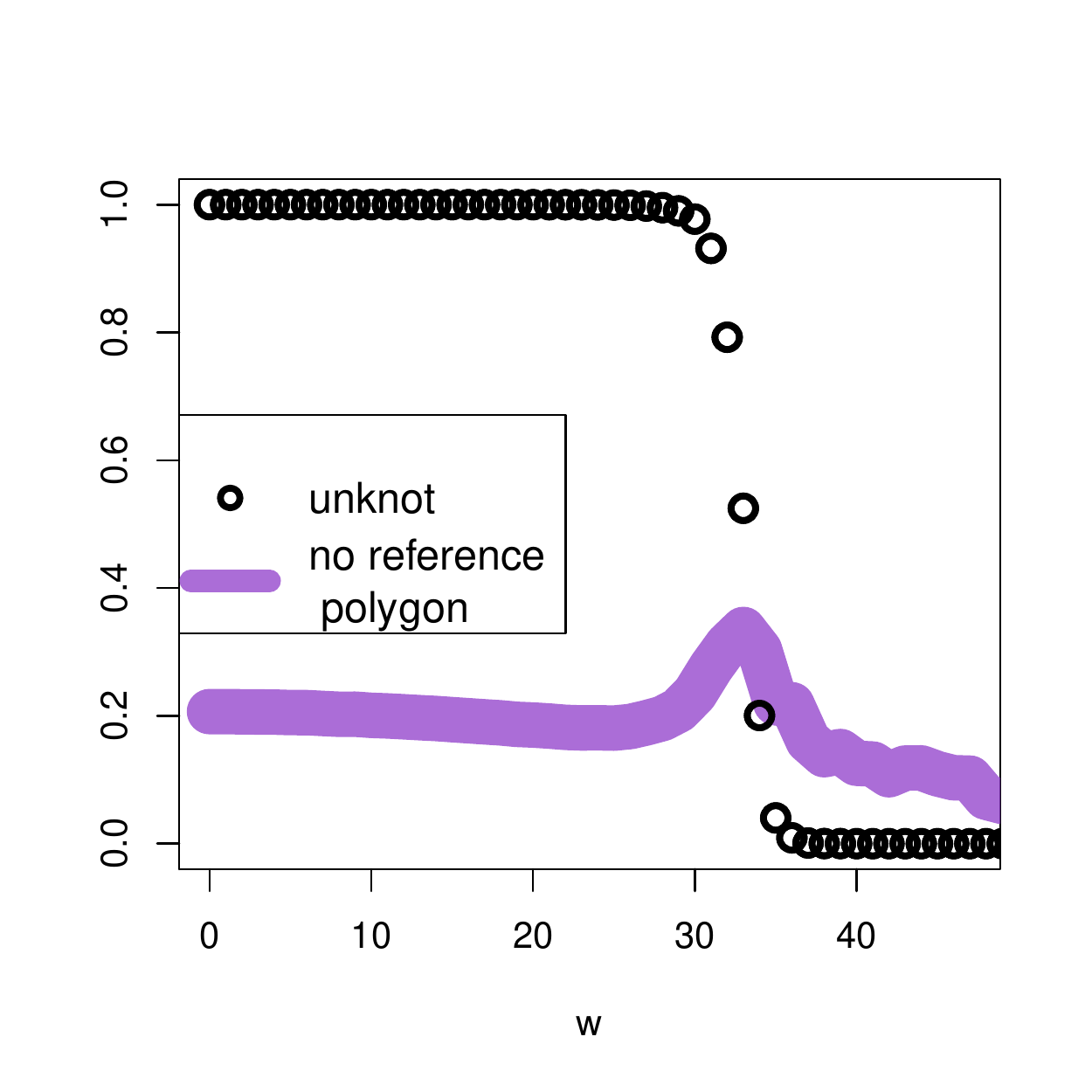}
\caption{\label{fig:unknot-density_N30}
For walks of length $n=24$, the figure shows the probability that walks with a fixed writhe $w$ have no reference polygon (thick/violet line).
It further shows the conditional probability that the reference polygon of these walks is an unknot, given that a reference polygon exists (open circles).}
\end{figure}

For walks of length $n=24$ that have a reference polygon, Figure~\ref{fig:unknot-density_N30} shows the probability that the reference polygon of walks with fixed
writhe $w$ is the unknot. There is a sharp transition between walks of small writhe where that probability is close to one and walks of large writhe where that probability is close to zero. Walks with $w<30$ are typically unknotted, whereas walks with writhe $w>36$ are typically knotted.
Figure~\ref{fig:unknot-density_N30} also shows the probability that walks of length $n=24$ with fixed writhe $w$ have no reference polygon. One can see that while
these walks exist for all values of $w$, the proportion is around $20\%$, and in fact decreases for large writhe. We thus conclude that in neither phase configurations without reference polygon are significant. We note that there is a peak in this probability near the transition, which is due to the fact that in configurations that are not tight, a reversal of the final step of the walk both increases the writhe and decreases the probability of the existence of an associated reference knot.
Configurations with large writhe are tight and typically have the ends of the walk on their exterior boundary, hence increasing the probability of the presence an associated reference polygon.

\section{Conclusion}

In this paper we considered the ensemble of self-avoiding walks in
the half-space on the simple cubic lattice, weighted by their writhe
and the distance of their endpoint from the surface. We showed that
there is a first-order phase transition between states of small and
high average writhe upon varying the torque. 

Our investigation of short lengths indicates that the low
and high writhe states are dominated by different  knot types.  More generally, this transition should be regarded as a transition between phases dominated by configurations of different knot type distributions. For
small torque, a SAW can contain a knot, with the knot type drawn from
a certain distribution of knot types. This clearly depends on the
length of the SAW; short SAWs are unknotted, whereas SAWs are almost
certainly knotted if they are sufficiently long. When increasing torque,
the distribution of knot types seem to change abruptly at a critical
value of the torque. This happens regardless of the strength of the
pulling force applied to the endpoint. Because this transition is
driven by the change of knot-type, we believe that it is insensitive
to the choice of lattice. 

Walks with less than 17 steps cannot knot, and the transition is absent.
In contrast, for a reasonably short walk of length $24$, we find
that below the transition the ensemble is dominated by unknotted configurations,
whereas above the transition the ensemble is dominated by trefoil
configurations having knot type $3_1$. 

In our simulations we have considered SAWs up to 256 vertices, which
when considering knot types for SAW is relatively short \cite{4472730720091021}. 
We cannot rule out the appearance of further phase transitions as
the length of the walk is increased, but it is certainly likely that
there exists at least one phase transition in the thermodynamic limit.
We further note that there are many composite knots with small writhe
\cite{1107.216220110711}, so that increasing the torque will bias
towards the appearance of certain types of knots.

Lastly, we have performed some short simulations weighting the linking number of lattice ribbons \cite{Thesis}.
These simulations indicate the same scenario as described here, with a transition associated with different knot types,
rather than the formation of plectonemes.

\section*{Acknowledgments}

One of the authors, ED, gratefully acknowledges the financial support
of the University of Melbourne via its Melbourne International Research
Scholarships scheme. Financial support from the Australian Research
Council via its support for the Centre of Excellence for Mathematics
and Statistics of Complex Systems and the Discovery Projects scheme
is gratefully acknowledged by one of the authors, ALO, who also 
thanks the School of Mathematical Sciences, Queen Mary University of London 
for hospitality. We also acknowledge support from VLSCI HPC and Edward HPC for providing computational resources.

\end{document}